\documentclass[aps,prl,twocolumn,superscriptaddress,showpacs]{revtex4-1}
\usepackage{graphicx}
\usepackage{subfigure}

\newcommand{\ket}[1] {| #1 \rangle}

\begin{document}

\title{The electron-phonon processes of the nitrogen-vacancy center in diamond}

\author{Taras Plakhotnik}
\email{taras@physics.uq.edu.au}
\affiliation{School of Mathematics and Physics, The University of Queensland, St Lucia, QLD 4072, Australia}

\author{Marcus W. Doherty}
\author{Neil B. Manson}
\affiliation{Laser Physics Centre, Research School of Physics and Engineering, Australian National University, ACT 2601, Australia}

\date{\today}

\begin{abstract}
Applications of negatively charged nitrogen-vacancy center in diamond exploit the center's unique optical and spin properties, which at ambient temperature, are predominately governed by electron-phonon interactions. Here, we investigate these interactions at ambient and elevated temperatures by observing the motional narrowing of the center's excited state spin resonances. We determine that the center's Jahn-Teller dynamics are much slower than currently believed and identify the vital role of symmetric phonon modes. Our results have pronounced implications for center's diverse applications (including  quantum technology) and for understanding its fundamental properties.
\end{abstract}


\maketitle

The negatively charged nitrogen-vacancy (NV) center is a point defect in diamond \cite{review} that has found diverse applications in quantum technology. The center is employed as a highly sensitive nanoscale sensor of electromagnetic fields \cite{mag1,mag2,mag3,mag4,efield1,efield2}, temperature \cite{toyli12,toyli13,neumann13,kucsko13,doherty14a,plakhotnik14} and pressure \cite{doherty14b} that can operate in ambient and extreme conditions. Recent NV metrology proposals include gyroscopy \cite{maclaurin12,ledbetter12,ajoy12,doherty14c} and the development of hybrid \cite{cai14} and multi-mode \cite{plakhotnik14} sensors.  In quantum information science, the NV center is used to realize spin registers \cite{qip1,qip2,qip3} at room temperature and spin-photon entanglement \cite{qip4,qip5} at cryogenic temperatures. A new direction in NV quantum information science seeks to exploit spin-phonon coupling to enhance NV spin registers and develop novel quantum devices \cite{mech1,mech2,mech3,mech4}.

The applications of the NV center are based upon its remarkable optical and spin properties. The center's room temperature applications primarily rely upon its bright optical fluorescence, long-lived ground state spin coherence and methods of optical spin polarization and readout. The latter enable the optical detection of the center's magnetic resonances (ODMR) and are the consequence of spin-dependent phonon-mediated intersystem crossings (ISCs) \cite{goldman1,goldman2}. The center's cryogenic applications also employ the coherence of the center's visible zero-phonon line (ZPL). The necessity of cryogenics arises from the temperature dependent electron-phonon induced dephasing and depolarization of the ZPL \cite{davies74,fu09}. Electron-phonon coupling is also responsible for the motional narrowing of the center's excited state spin resonances, which determines their utility as an additional quantum resource for sensing and information processing \cite{plakhotnik14,fuchs10}. Thus, a through understanding of the NV center's electron-phonon interactions is important to the continued advancement of its applications and may be generalized to similar defects with emerging quantum applications, such as the silicon-vacancy center in diamond \cite{siv1,siv2} and centers in silicon carbide \cite{sic1,sic2}. Here, we show that there exist several issues in the current understanding and identify possible resolutions.

\begin{figure}
\begin{center}
\includegraphics[width=0.65\columnwidth]{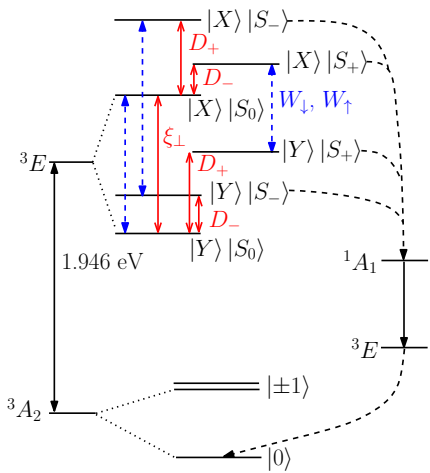}
\caption{\label{Fig1} The electronic and fine structures of the NV center at high stress. The $^3E$ sub-levels are labelled by their product of orbital ($\ket{X}$, $\ket{Y}$) and spin ($\ket{0}$, $\ket{S_\pm}$) states, where the spin states are solutions of (\ref{spinhamiltonian}). The $^3E$ fine structure splittings ($D_\pm=D_\parallel\pm D_\perp$) are denoted in red. The $^3A_2$ sub-levels are denoted by their spin projection ($m_s=0$,$\pm1$). The optical transitions of the spin triplet and singlet levels are depicted as black solid arrows and the visible ZPL is at $1.956$ eV. Blue dashed arrows represent the population transfers within the $^3E$ (rates $W_\downarrow$ and $W_\uparrow$). The black dashed arrows denote the allowed ISCs between the spin triplet and singlet levels.}
\end{center}
\end{figure}

The electronic structure of the NV center is depicted in Fig. \ref{Fig1}. The optical transitions of the visible ZPL occur between the ground $^3A_2$ and excited $^3E$ spin triplet levels. The temperature dependent broadening of the ZPL was initially described \cite{davies74} using the widely applicable model of quadratic electron-phonon interactions with $A_1$ phonon modes \cite{maradudin66}. However, subsequent single center cryogenic measurements revealed that the broadening was more consistent with the characteristic $\propto T^5$ temperature dependence of linear electron-phonon (Jahn-Teller) interactions with $E$ phonon modes \cite{fu09}. These interactions induce population transfer between the quasi-degenerate orbital states ($\ket{X}$, $\ket{Y}$) of the $^3E$ level (see Fig. \ref{Fig1}), which dephases the optical transitions and leads to the depolarization of the ZPL fluorescence \cite{fu09}. Applying their Jahn-Teller model, Fu et al \cite{fu09} identified a factor of $\sim2$ inconsistency between the population transfer rates that describe the ZPL broadening and depolarization at low temperatures. By introducing a phonon cutoff energy, Abtew et al \cite{abtew11} attempted to extend Fu et al's model to describe the ZPL broadening up to room temperature. In doing so, they obtained a cutoff at 50 meV for $E$ phonons, which is much lower than the diamond Debye energy $\omega_D\approx$168 meV \cite{zaitsev} and the features of the NV electron-phonon spectral density extracted from the visible phonon sideband \cite{kehayias13}.

At  temperatures $\lesssim30$ K, the complicated six level $^3E$ fine structure (see Fig. \ref{Fig1}) is observed via high resolution optical spectroscopy \cite{goldman1}. Above $\approx150$ K \cite{batalov09}, the population transfer between the $^3E$ orbital states is sufficiently fast to dynamically average the $^3E$ fine structure so that ODMR resembles the simpler three level structure of the ground state \cite{rogers09}. The dynamically averaged fine structure is temperature dependent and is described by the spin Hamiltonian \cite{plakhotnik14}
\begin{eqnarray}
H & = & D_\parallel (S_z^2-\frac{2}{3})-D_\perp R(T)(S_x^2-S_y^2)
\label{spinhamiltonian}
\end{eqnarray}
where $D_\parallel=1.42$ GHz and $D_\perp=0.775$ GHz are the $^3E$ spin-spin parameters, $R(T)=(e^{h\xi_\perp/k_B T}-1)/(e^{h\xi_\perp/k_B T}+1)$ is the temperature reduction factor, $h$ and $k_B$ are the Planck and Boltzman constants, respectively, and $\xi_\perp$ is the $^3E$ strain splitting. Note that the negligible contribution of the $\lambda_\perp$ spin-orbit term (see Ref. \onlinecite{plakhotnik14}) to the fine structure is ignored here.

The dynamical averaging is also expected to motionally narrow the $^3E$ ODMR, since the rapid population transfer decouples the orbit and spin degrees of freedom. Fuchs et al have measured the $^3E$ spin dephasing rate at room temperature \cite{fuchs10}. They attributed the observed dephasing to the dynamical averaging process and, using a motional narrowing model, suggested that elevated temperatures or strain may decrease its rate. However, their proposal is yet to be tested by a systematic study of the motional narrowing effect. Furthermore, there is an inconsistency between Fuchs et al's observations and the current ZPL broadening model. If the population transfer rate ($\sim 10$ THz) at room temperature is inferred from the ZPL width \cite{fuchs10}, then the spin dephasing rate predicted by the motional narrowing model ($\sim1.2$ MHz) is almost two orders of magnitude smaller than measured ($\sim92$ MHz). The anomalously low cutoff of Abtew et al, the discrepancy identified by Fu et al, the conspicuous absence of interactions with $A_1$ modes, and the orders of magnitude larger than expected $^3E$ spin dephasing rate, all indicate problems in the current ZPL broadening model.

The optical polarization and readout of the spin triplet levels is a result of spin-selective ISCs with the intermediate $^1A_1$ and $^1E$ spin-singlet levels (see Fig. \ref{Fig1}) \cite{review}. Goldman et al \cite{goldman1,goldman2} have developed a detailed model of the electron-phonon mechanisms that govern the ISC from $^3E$ to $^1A_1$. However, in order to validate Goldman et al's model and extend it to room temperature, more detailed and quantitative knowledge of linear $E$ phonon interactions is required. This knowledge can be improved by extending the current measurements of the population transfer rates at cryogenic temperatures to room temperature and beyond.

In this paper, we report observations of the $^3E$ ODMR of NV centers in nanodiamond over the temperature range 295-550 K. We show that the ODMR is well described by a motional narrowing model and extract the population transfer rates. We establish that the rates are much slower than currently believed and do not account for the observed ZPL broadening at room temperature. We propose that quadratic $A_1$ phonon interactions contribute significantly to the ZPL already above 30 K. Finally, we back up and rectify the proposals of Fuchs et al, resolve the inconsistencies of the ZPL broadening model and provide valuable insight into electron-phonon coupling above cryogenic temperatures.

\begin{figure}
\begin{center}
\includegraphics[width=0.9\columnwidth]{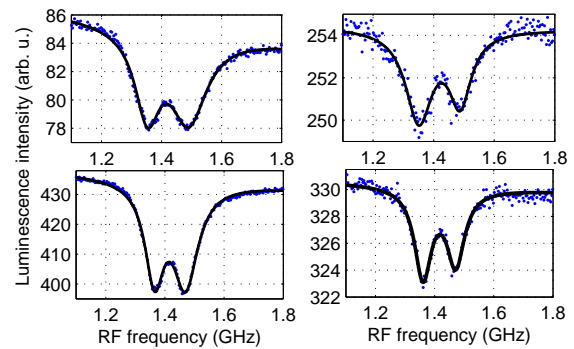}
\caption{\label{Fig2} Example ODMR spectra at different temperatures (315 K upper, 455 K lower) and RF powers (440 mW left, 55 mW right). The narrowing and reduced splitting of the lines at higher temperature as well as power broadening at higher RF power can be seen. The lineshape fits (solid lines) are the sum of two Lorentzians and a linear background.}
\end{center}
\end{figure}

Our continuous wave ODMR experiments were performed using 532 nm laser excitation and fluorescence collection via an epifluorescence design. The nanodiamonds were spin coated on a silica substrate. The NV spin resonances were driven by radio-frequency  (RF) magnetic field created by a gold wire deposited onto the substrate. The excitation laser spot overlapped with the wire and the optical heating of the wire was used to control the temperature of a chosen nanodiamond. See Ref. \onlinecite{plakhotnik14} for further experimental details. On average, the nanodiamonds had a diameter of $\sim30$ nm and contained $\sim15$ NV centers. We performed ODMR measurements on a total of 10 nanodiamonds. The results from one nanodiamond are presented here and are consistent with the rest of the sample and, as will be explained, measurements in bulk diamond. Previous optical spectroscopy has measured the $^3E$ strain splitting of the nanodiamond to be $h\xi_\perp\sim4.7$ meV \cite{plakhotnik14}. This large strain splitting permits a simple $^3E$ fine structure (see Fig. \ref{Fig1}) and application of the motional narrowing model. Note that the strain splitting in previous reports \cite{fu09,fuchs10} have been much smaller.

Examples of ODMR spectra are shown in Fig. \ref{Fig2}. Averaging over the unresolved $^3E$ hyperfine structure, the observed ODMR splitting is \cite{plakhotnik14}
\begin{eqnarray}
\Delta_\mathrm{ODMR} & = & \frac{2}{3}D_\perp R(T)+\frac{4}{3}\left[A^2+D_\perp^2 R^2(T)\right]^{1/2},
\end{eqnarray}
where $A\approx40$ MHz is the isotropic hyperfine parameter. RF-power broadening is evident in Fig. \ref{Fig2}. Similar to the analysis of the $^3A_2$ ODMR in Ref. \onlinecite{jensen13}, a five-level model of the optical and spin dynamics yields the following expressions for the ODMR linewidth $\Gamma_\mathrm{ODMR}$ and contrast $C_\mathrm{ODMR}$
\begin{eqnarray}
\Gamma_\mathrm{ODMR} & = &\Gamma_\mathrm{ODMR}^\mathrm{(inh)} +\Gamma_\mathrm{ODMR}^\mathrm{(h)} \left(1+\frac{4\pi \kappa P_\mathrm{RF} }{\Gamma_\mathrm{ODMR}^\mathrm{(h)} \gamma _{1} } \right)^{1/2} \nonumber \\
C_\mathrm{ODMR} & = &C_\mathrm{ODMR}^{{\rm (max)}} \frac{4\pi \kappa P_\mathrm{RF} }{4\pi \kappa P_\mathrm{RF} +\gamma _{1} \Gamma_\mathrm{ODMR}^\mathrm{(h)}},
\label{Gamma_sat}
\end{eqnarray}
where $\Gamma_\mathrm{ODMR}^\mathrm{(h)}$ and $\Gamma_\mathrm{ODMR}^\mathrm{(inh)}$ are the homogenous and inhomogenous linewidths in the absence of power broadening, $P_\mathrm{RF}$ is the RF-power, $\kappa$ is a proportionality factor such that $\kappa P_\mathrm{RF}$ is the spin Rabi frequency, and $\gamma_1$ is the effective spin relaxation rate. The essential difference to Ref. \onlinecite{jensen13} is a much weaker, but more complicated dependence of $\gamma _{1} $ on the laser power.  At low laser powers, $\gamma _{1} \approx k k_\mathrm{ISC}/(k + 0.5k_\mathrm{ISC})\approx 22$ MHz, where $k\approx20$ MHz \cite{taras2012} is the $^3E$ radiative decay rate in nanodiamond of the same type and origin as used in this work and $k_\mathrm{ISC}\approx50$ MHz is the average $^3E$ ISC rate \cite{goldman1,taras2012}. Stress inhomogeneity and the unresolved hyperfine structure contribute to $\Gamma_\mathrm{ODMR}^\mathrm{(inh)}$.

The homogenous linewidth $\Gamma_\mathrm{ODMR}^\mathrm{(h)}=\Gamma_\infty+\Gamma_\mathrm{MN}(T)$ is the sum of the broadening due to the $^3E$ orbital decay rate $\Gamma_\infty =(k+0.5k_\mathrm{ISC})/\pi$ and motional narrowing $\Gamma_\mathrm{MN}(T)$. Whilst the orbital decay rate increases at high temperature \cite{toyli12,plakhotnik10}, this temperature dependence is ignored in the following because the contribution of $\Gamma_\infty$ to the observed  $\Gamma_\mathrm{ODMR}$  changes little, from 14 MHz to 17 MHz between 295 K and  500 K. The spin bath dephasing contribution to $\Gamma_\infty$ is also ignored because it has been assessed using the $^3A_2$ ODMR to be negligibly small (1-2 MHz). In the fast exchange approximation of motional narrowing \cite{reilly94,slitcher}, where the population transfer rates ($W_\uparrow$, $W_\downarrow$) are much larger than the jump in the spin resonances between the $^3E$ orbital states ($2D_\perp$),
$\Gamma_\mathrm{MN}(T) \approx  \beta(T)2\pi D_\perp^2/W_\downarrow$. The factor $\beta(T)=8e^{-h\xi_\perp/k_B T}/(e^{-h\xi_\perp/k_BT}+1)^3$ is close to 1 above room temperatures. Thus, as $W_\downarrow$ increases with temperature, $\Gamma_\mathrm{MN}$ decreases.

In the temperature regime $k_B T \gg h\xi_\perp$, Raman scattering of $E$ phonons dominate the population transfer rates which read \cite{goldman1}
\begin{eqnarray}
W_\downarrow & = B_ET^5\int_{x_\perp}^{\frac{\Omega_E}{k_B T}}\frac{x^2e^x(x-x_\perp)^2}{(e^x-1)(e^{x-x_\perp}-1)}dx
\label{eq:wdown}
\end{eqnarray}
 and $W_\uparrow=W_\downarrow e^{-h\xi_\perp/k_B T}$, where $x_\perp=h\xi_\perp/k_B T$ and $\Omega_E$ is the cutoff energy for $E$ phonons. The deformation potential and Debye density of states for acoustic phonons have been assumed such that the corresponding electron-phonon spectral density is $J_E(\omega)\approx\eta_E\omega^3$ and the constant $B_E=\frac{64}{\pi}\hbar\eta_E^2k_B^5$.   Whilst in the simplest case $\Omega_E=\omega_D$, the cutoff is often considered as a phenomenological parameter which takes into account the departure from $J_E(\omega) \propto\omega^3$. In the high temperature regime $1 \gg x_\perp, \hbar\Omega_E/k_B T$ applicable to our work, the integral above equals $\frac{1}{3}\left(\frac{\Omega_E}{k_BT}\right)^3\left(1-\frac{h\xi_\perp}{\Omega_E}\right)^2\left(1+\frac{h\xi_\perp}{2\Omega_E}\right)$ and  $W_\downarrow = QT^2$, where $Q$ is a constant. Hence, we obtain the final expression $\Gamma_\mathrm{ODMR}^\mathrm{(h)}=\Gamma_\infty+\beta(T)2\pi D_\perp^2/(QT^2)$.

\begin{figure}
\begin{center}
\includegraphics[width=1\columnwidth]{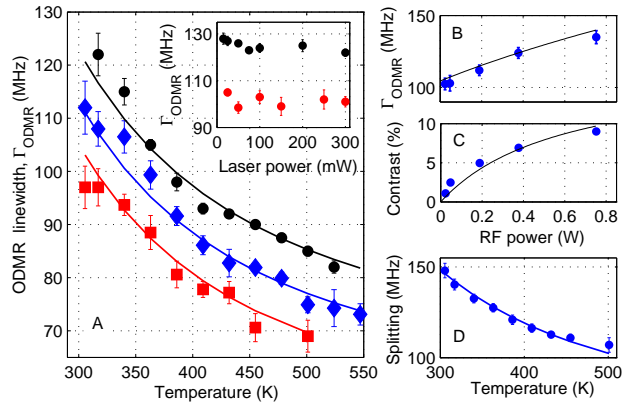}
\caption{\label{Fig3} A: ODMR linewidth as a function of temperature at RF-powers of 400, 200, and 50 mW (top to bottom). Inset shows the weak optical power dependence of the linewidth at 294 K and at two RF powers: 47 mW (bottom) and 380 mW (top). B and C show the RF-power dependence of the linewidth and contrast at 294 K and 100 mW optical power. D: ODMR splitting at different temperatures (50 mW RF-power). Error bars are determined by the statistics of repeated measurements. The plotted linewidth is the average width of the two lines.}
\end{center}
\end{figure}

Systematic measurements of the ODMR linewidth, contrast and splitting at different temperatures, RF and laser powers are presented in Fig. \ref{Fig3}. The weak optical-power dependence [inset of Fig. \ref{Fig3}(a)]  supports the approximation $\gamma_1\approx 22$ MHz. The simultaneous fitting of the six data sets using the five parameters yields $\Gamma_{{\rm ODMR}}^\mathrm{(inh)} =33\pm 3{\rm \; MHz}$, $\kappa \approx 210\pm 40{\rm \; MHz}^{{\rm 2}} {\rm \; W}^{{\rm -1}} $, $C_{{\rm ODMR}}^{{\rm (max)}} =16\pm 2\% $, $Q=0.83\pm 0.06{\rm \; MHz \;K}^{{\rm -2}} $, and $h\xi_\perp =4.6\pm 0.2$ meV. The values of $\kappa$ and $h\xi_\perp$ are in reasonable agreement with the parameters of the RF wire and previous optical spectroscopy, respectively. The fitting yields $\Gamma_\mathrm{ODMR}^\mathrm{(h)}=55$ MHz at room temperature. The dephasing rate measured by Fuchs et al at room temperature in bulk diamond corresponds to $\Gamma_\mathrm{ODMR}^\mathrm{(h)}\sim29$ MHz. Taking into account that the much smaller stress splitting $\xi_\perp$ of Fuchs et al's NV center will increase $Q$ by $\sim2$, the two values are in agreement. Hence, we conclude that our nanodiamond measurements are consistent with bulk diamond and capture intrinsic phenomena of the NV center.

\begin{figure}
\begin{center}
\includegraphics[width=1\columnwidth]{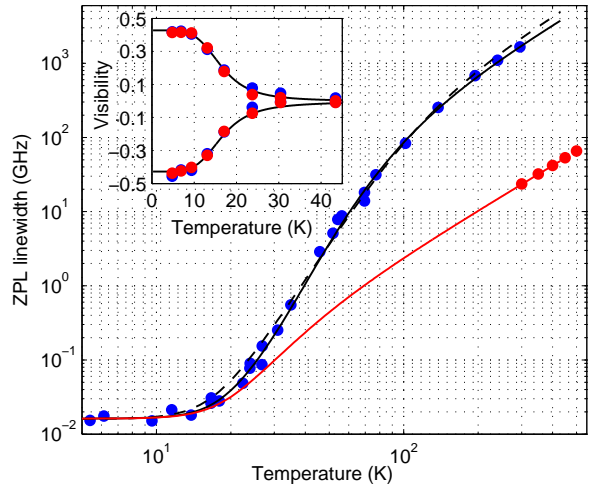}
\caption{\label{Fig4}  Blue points are the ZPL width measured in Ref. \onlinecite{fu09}. The black solid curve depicts the fit of our model and the black dashed curve is the extended Jahn-Teller model of Ref. \onlinecite{abtew11}. The red solid curve is the contribution of $W_{\downarrow }$ to the ZPL width according to (\ref{eq:wdown},\ref{eq:zplmodel}). The red dots show $W_{\downarrow }/2\pi$ derived from ODMR data alone. Inset: the ZPL polarization visibility of two NV centers (red and blue points) from \cite{fu09}. The solid curve is our fit obtained using the model $V=\left(W_{\uparrow } -W_{\downarrow } \pm r\left(1-a\right)/\left(1+a\right)\right)/\left(W_{\downarrow } +W_{\uparrow } +r\right)$, where $a=0.40\pm 0.02$ and $r=80{\rm \; MHz}$ are defined in Ref. \onlinecite{fu09}, and $W_\downarrow$ and $W_\uparrow$ are determined by our fit of the ZPL width.}
\end{center}
\end{figure}

The previous measurements of the ZPL width \cite{fu09} are plotted in Fig. \ref{Fig4} together with $W_\downarrow /(2\pi)$ calculated here using the value of $Q$ that we obtained by fitting our motional narrowing observations (rescaled to $\xi_\perp=0$ to match the stress splitting in Ref. \onlinecite{fu09}). It is evident that the rates are orders of magnitude too small to account for the ZPL width alone. We propose that the additional width is due to quadratic interactions with $A_1$ modes that purely dephase the optical transitions \cite{davies74,maradudin66}. In which case, the ZPL width is \cite{cohen}
\begin{eqnarray}
\Gamma_\mathrm{ZPL}=\frac{W_\downarrow}{2\pi}+\frac{W_A}{\pi}+\gamma_0,
\label{eq:zplmodel}
\end{eqnarray}
where $W_A$ is the additional dephasing rate and $\gamma_0$ is the approximately temperature independent contribution of the optical decay rate. As per a similar derivation of $W_\downarrow$,
\begin{eqnarray}
W_A & = & B_A T^7 \int_0^{\frac{\Omega_A}{k_BT}} \frac{e^xx^6}{(e^x-1)^2}dx,
\end{eqnarray}
where $B_A$ is a constant and $\Omega_A$ is the cutoff energy of $A_1$ phonons.  We used Eqs.(4-6) and fitted the ZPL width measurements to obtain $B_E=1.32$ Hz K$^{-5}$, $\Omega_E=13\pm1$ meV,  $B_A=24\pm4\; \mu$Hz K$^{-7}$, $\Omega_A=37\pm2$ meV and $\gamma_0=16.2\pm0.5$ MHz (in bulk diamond). We confirmed our parameters ($B_E$ and $\Omega_E$) of the population transfer rates by also fitting the polarization visibility measurements of Ref. \onlinecite{fu09} (see Fig. \ref{Fig4}). Our fit of the visibility curve is practically indistinguishable from Ref. \onlinecite{fu09} and our value of $B_E$ also agrees with the value $\sim1.6$ Hz K$^{-5}$ obtained there. Unlike  Ref. \onlinecite{fu09}, we use the same value of $B_E$  for ZPL and visibility fits and our fit to ZPL data  better describes the low and room temperature regions than the extended Jahn-Teller model presented in Ref. \onlinecite{abtew11}. Most importantly, the ZPL broadening is fully consistent with our ODMR measurements at elevated temperatures.

The phonon cutoffs that we obtained are much lower than expected. We attribute this to the inadequacies of the acoustic approximation of the phonon spectral density $J(\omega)\approx\eta_E\omega^3$ \cite{davies74,maradudin66} and consider the cutoffs as phenomenological. Noting that $\Omega_E/k_B\sim155$ K, these inadequacies are negligible at low temperatures, which explains why they were not detected in previous cryogenic measurements \cite{fu09,goldman1}. Interestingly, $\Omega_E$ is close to the calculated Jahn-Teller barrier energy $\sim10$ meV \cite{abtew11}. Note that the spectral density extracted, for example, from the visible phonon sideband represents contributions of  $E$ and $A_1$ phonons due to linear electron-phonon interactions. It is difficult to  distinguished the  effects of $A_1$ and $E$ modes on the phonon band experimentally and \textit{ab initio} calculations therefore appear to be the best avenue for future advancement to resolve the puzzle.

\begin{acknowledgments}
This work was supported by the Australian Research Council under the Discovery Project scheme DP0771676 and DP120102232.
\end{acknowledgments}

\end{document}